\documentclass[apr,prl,twocolumn,superscriptaddress,super,floatfix]{revtex4-2}
\usepackage[utf8]{inputenc}
\usepackage[T1]{fontenc}
\usepackage[english]{babel}
\usepackage{amsmath}
\usepackage{braket}
\usepackage{bm}
\usepackage{chemformula}
\usepackage{siunitx}
\usepackage{amsbsy}
\usepackage{float}
\usepackage[version=4]{mhchem}

\usepackage{color}
\usepackage{graphicx}
\usepackage{amssymb}
\usepackage{amsfonts,dsfont}
\usepackage{subfigure}

\usepackage[normalem]{ulem}
\usepackage[bbgreekl]{mathbbol}
\usepackage{mathrsfs,empheq}
\usepackage{relsize,scalerel}

\usepackage[toc,page]{appendix}

\usepackage[savepos]{zref}

\usepackage{siunitx}[=v2]

\newcommand{\mrm}[1]{\mathrm{#1}}

\renewcommand{\sout}[1]{}

\begin{document}

\pdfpagewidth=8.5in
\pdfpageheight=11in
\pdfpageattr{/Rotate 0}

\title{Displacive quantum critical point in superconducting hydrides: The case of $\ce{H_3S}$}

\author{Marco Cherubini}
\email{marco.cherubini@sorbonne-universite.fr}
\affiliation{Institut de Minéralogie, de Physique des Matériaux et de Cosmochimie, Sorbonne Université, CNRS UMR 7590, MNHN, 4 Place Jussieu, Paris, 75005, France}

\author{Abhishek Raghav}
\affiliation{Institut de Minéralogie, de Physique des Matériaux et de Cosmochimie, Sorbonne Université, CNRS UMR 7590, MNHN, 4 Place Jussieu, Paris, 75005, France}
\affiliation{RIKEN Center for Emergent Matter Science, 2-1 Hirosawa, Wako-shi, Saitama 351-0198, Japan}

\author{Michele Casula}
\email{michele.casula@sorbonne-universite.fr}
\affiliation{Institut de Minéralogie, de Physique des Matériaux et de Cosmochimie, Sorbonne Université, CNRS UMR 7590, MNHN, 4 Place Jussieu, Paris, 75005, France}

\date{\today}

\begin{abstract}
H$_3$S sulfur hydride has been widely investigated for its high superconducting critical temperature $T_c$ of 203 K at about $p_c = 155$ GPa. Despite being the precursor of superconducting hydrides, a detailed picture of its structural phase diagram in an extended temperature and pressure range is still missing. To determine it with 
inclusion of both thermal and quantum effects, we carry out path integral molecular dynamics 
combined to a MACE neural network potential trained on BLYP density functional theory configurations. The resulting H$_3$S phase diagram is characterized by the displacive transition between the centrosymmetric Im$\bar{3}$m and polar R3m phases, which originates from a quantum critical point (QCP) located at $p_\mrm{QCP} \approx 134$ GPa. We show that the experimental $T_c$ peak falls into a centrosymmetric region of large nuclear quantum fluctuations above the displacive QCP, as measured by local phonon Green's functions resolved in imaginary time, where fluctuating moments are at play. We study the critical behavior of the system in the proximity of the QCP by a finite-size scaling analysis, showing that it belongs to the 4D Ising universality class. We finally discuss its implications for the superconducting state.
\end{abstract}

\maketitle

Superconductivity was first discovered more than a century ago\cite{Onnes_supra}. Since then, the quest for systems with high superconducting critical temperature ($T_c$) has been one of the frontiers in both experimental and theoretical physics, 
with high values of $T_c$ expected in systems containing light atoms\cite{Ashcroft1968,Gorkov2016}. In 2015, a $T_c$ of 203 K at about 155 GPa was measured in sulfur hydride ($\ce{H_3S}$)\cite{Drozdov_2015,Einaga_2016,mozaffari_2019,Minkov_2020,Osmond_2022}, exceeding the highest values found in cuprates\cite{Bednorz1986,Gao1994,Bianconi2015}. In $\ce{H_3S}$, the superconducting critical temperature shows a dome-like shape as a function of pressure\cite{Drozdov_2015}, similarly to what happens in cuprates\cite{Keimer2015,Michon2019} or in certain dilute metals like $\ce{SrTiO_3}$\cite{Ngai1974,Kiselov2021,vanderMarel2019}. The $T_c$ peak at 155 GPa was originally associated with a displacive phase transition\cite{Minkov_2020} from the body-centered-cubic
Im$\bar{3}$m (centrosymmetric) structure to the lower-symmetry trigonal R3m (polar)  phase\cite{duan2014pressure,flores2016high,goncharov_2017}, triggered by the hydrogen (H) atoms displaced from their midpoint position between two flanking sulfur (S) atoms.
However, theoretical attempts to study this transition failed to locate it at a pressure compatible with the position of the $T_c$ peak, the most reliable theories placing the transition pressure more than 40 GPa below\cite{Errea2016,Bianco_2018,Taureau2024}.
Nevertheless, a detailed picture of the $\ce{H_3S}$ structural
phase diagram is still missing, particularly its temperature evolution, which remains largely unexplored.

\begin{figure}[b!]
    \centering
    \includegraphics[width=0.95\columnwidth]{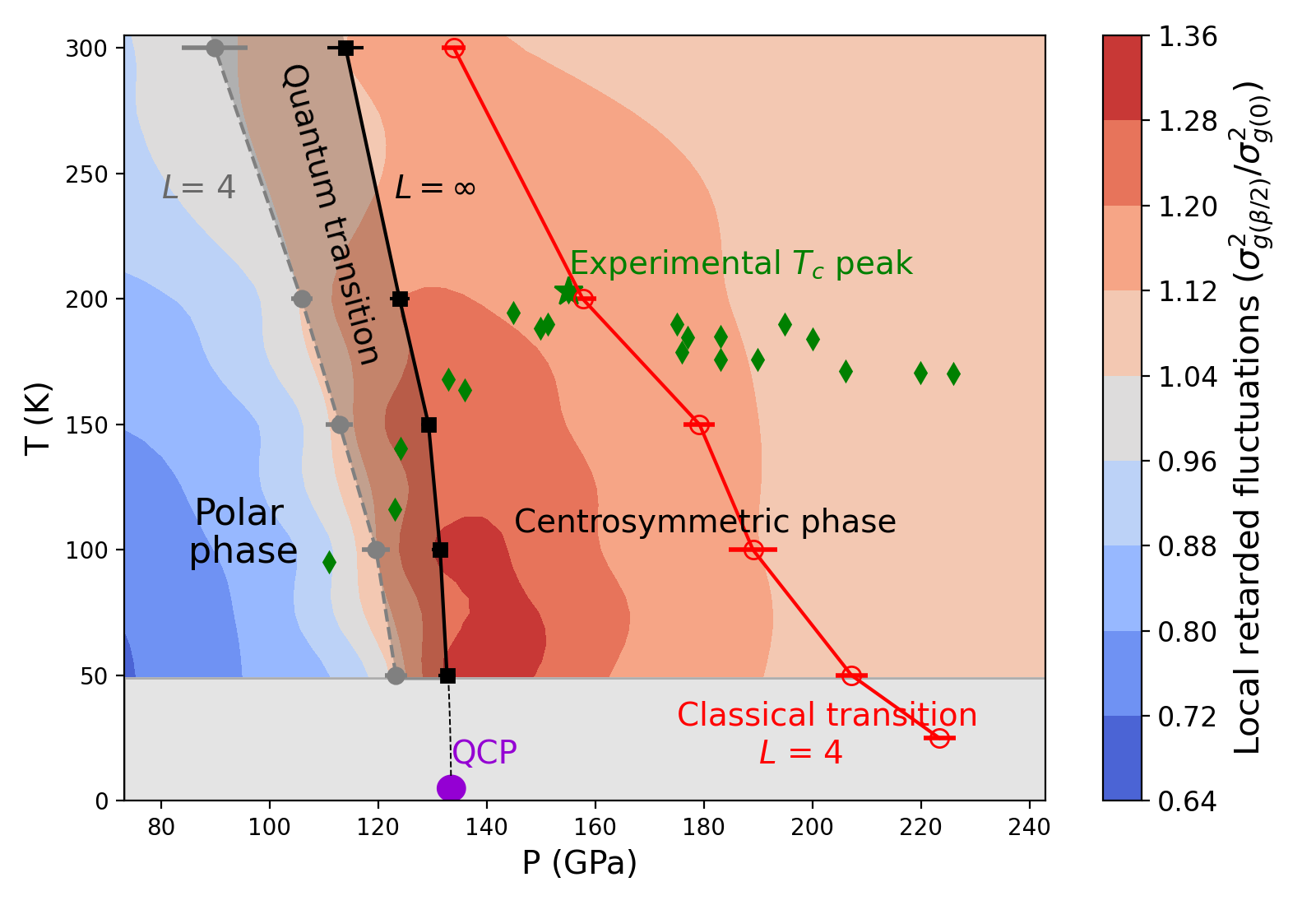}
    \caption{\ce{H_3S} phase diagram, 
    computed by PIMD using a MACE-MLIP trained with a DFT-BLYP dataset for $L=4$: displacive transition (gray points, dashed line); classical molecular dynamics (MD) transition (red empty circles, solid line); extrapolated QCP (filled purple circle) and displacive line (black squares, solid line); 
    experimental superconducting $T_c$ (green diamonds)\cite{Drozdov_2015,Einaga_2016}.
    The colormap refers to the $\sigma^2_{g(\beta/2)}/\sigma^2_{g(0)}$ ratio plotted in Fig.~\ref{fig:order_parameters}(d); it measures the relative intensity of the retarded fluctuations of local moments. Data are interpolated using a weighted sum of radially symmetric basis functions. Blue and red shaded areas are the polar and centrosymmetric phase, respectively. 
    }
    \label{fig:phase-diagram}
\end{figure}

In this Letter, we fill this gap by studying the system over a large temperature and pressure range
and by analyzing its quantum criticality in the proximity of the displacive transition. By
path integral molecular dynamics calculations (PIMD)
with machine learning interatomic potential (MLIP) trained on density functional theory (DFT-BLYP)\cite{Becke_1988,lee1988development},
we unveil the presence of a displacive quantum critical point (QCP) at 134 $\pm$ 2 GPa belonging to the 4D Ising universality class (Fig.~\ref{fig:phase-diagram}). 
DFT-BLYP yield accurate pressures in this range, as verified in Ref.~\onlinecite{Taureau2024} through a comparisons against benchmark quantum Monte Carlo calculations.
According to our picture, the dome shape of the $T_c$ evolution has a peak falling in the centrosymmetric Im$\bar{3}$m phase, and it is located in a region of strong quantum fluctuations surrounding the QCP, with superconductivity possibly enhanced by them.

\begin{figure*}[t!]
    \centering
    \includegraphics[width=0.8\textwidth]{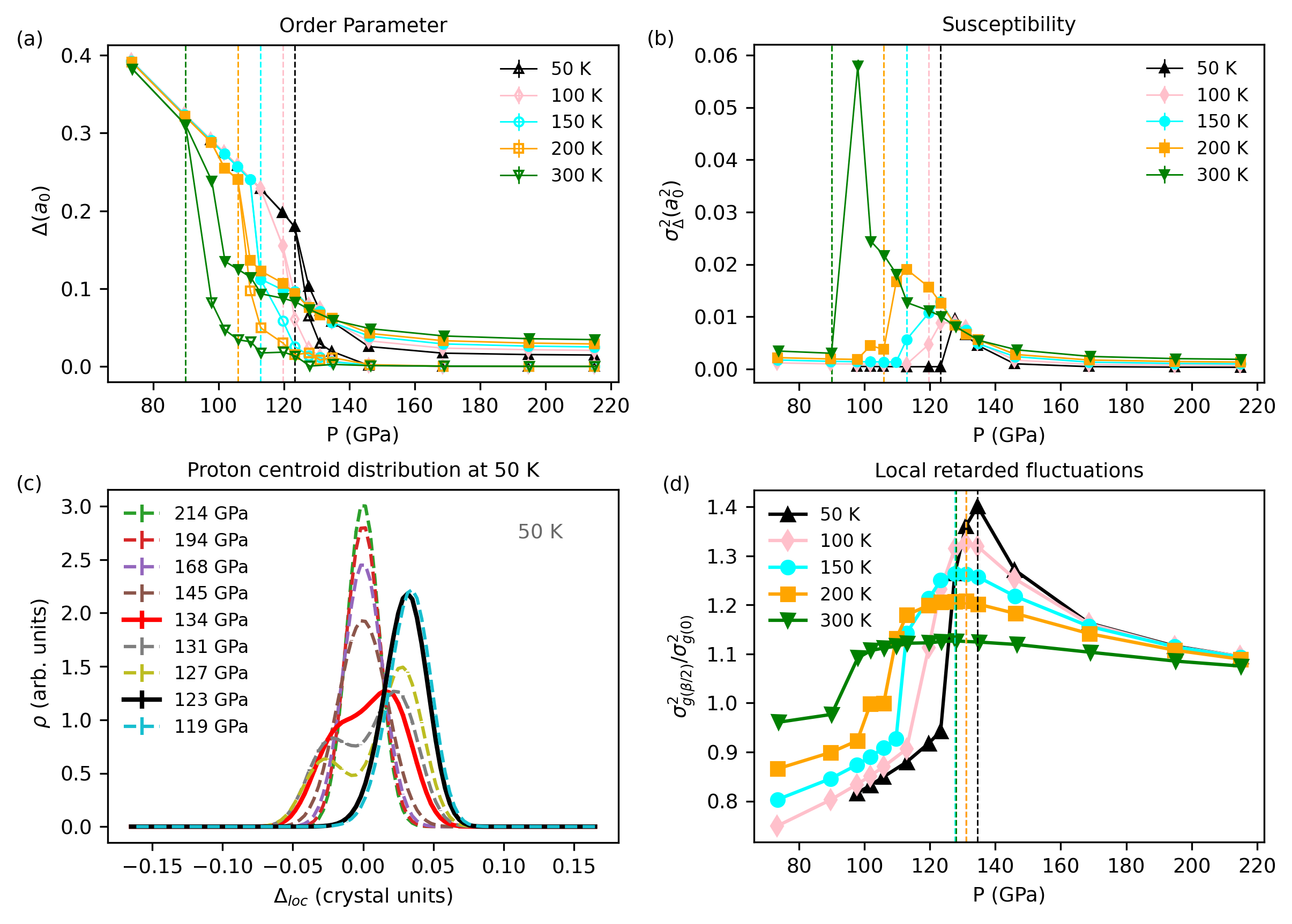}
    \caption{Order parameter and local observables for $L=4$. Panel (a): Order parameter for the displacive transition. Full and empty symbols are the absolute and the real values of the order parameter, $\Delta_{\textrm{abs}}$ and $\Delta$, respectively. The variance of $\Delta$ is in panel (b). Different temperatures are reported: 50 K (black triangles), 100 K (pink diamonds), 150 K (cyan circles), 200 K (orange squares) and 300 K (green reversed triangles). The vertical dashed lines indicate the transition pressures. Panel (c): Proton centroids distribution along the S-S direction for several pressures at $T=50$ K. Thick black and red curves indicate the displacive transition and the unimodal-to-bimodal variation, respectively. Panel (d): $\sigma^2_{g(\beta/2)}/\sigma^2_{g(0)}$ ratio, where $\sigma^2_{g(\tau)}$ is the variance of the local Green function in imaginary time. 
    Dashed lines indicate its maximum.
    }
    \label{fig:order_parameters}
\end{figure*}

In systems containing light atoms, like $\ce{H_3S}$, nuclear quantum fluctuations play a pivotal role. 
In this work, we use the PIMD formalism to treat nuclear quantum effects (NQE) exactly.
However, its computational cost increases
significantly as the temperature is lowered.
Therefore, the combination of PIMD with \emph{ab initio} electronic structure methods, such as DFT, to explore the full phase diagram of sulfur hydride is unfeasible. Here, to maintain the DFT-level accuracy at a much lower computational cost, we trained a
MLIP using MACE\cite{Batatia2022mace,Batatia2022Design}. Then, we ran NVT-PIMD simulations at different volumes $V$ and temperatures $T=1/(k_B \beta)$, using the i-PI package\cite{Ceriotti2014} in combination with the generated MLIP. Quantum nuclei are necklaces made of $L_\tau$ beads, such that $\beta= L_\tau \, \delta\tau$, with $\delta\tau$ the imaginary time step assuring convergence.
The simulations are up to 400 ps long, for systems made of $L\times L\times L$ primitive cells with $L \in \{2, 3, 4\}$, necessary for an accurate determination of phase boundaries via finite-size scaling. 
Further details about the PIMD simulations and the MLIP generation can be found in the Supplemental Material (SM)\cite{SM}. 

The Im$\bar{3}$m and R3m structures differ by
the average protons position with respect to their neighboring S atoms. 
To distinguish the two symmetries,  we define the global proton displacement
\begin{equation}
    \label{eq:order_parameter}
    \Delta^{(j)} = \frac{1}{N_\mrm{H}}\sum_{i=1}^{N_\mrm{H}} \bigl[ \vec{r}^{(j)}_{\mrm{H}_i\mrm{S}_{i_1}} \cdot \hat{r}^{(j)}_{\mrm{S}_{i_1}\mrm{S}_{i_2}} - d^{(j)}_{\mrm{S}_{i_1}\mrm{S}_{i_2}}/2],
\end{equation}
evaluated at the $j$-th PIMD iteration. For 
every hydrogen atom $\mrm{H}_i$ in the supercell, we identify its two flanking S
atoms, $\mrm{S}_{i_1}$ and $\mrm{S}_{i_2}$. Then, we project the 
vector $\vec{r}_{\mrm{H}_i\mrm{S}_{i_1}}$, connecting $\mrm{H}_i$ to one of the two S atoms, onto 
the $\mrm{S}_{i_1}$-$\mrm{S}_{i_2}$ direction.
Finally, the distance 
between the projection and the $\mrm{S}_{i_1}$-$\mrm{S}_{i_2}$
midpoint is summed over the $N_\mrm{H}$ hydrogen atoms in the supercell.
$\mrm{H}_i$, $\mrm{S}_{i_1}$ and $\mrm{S}_{i_2}$ are taken here as centroid positions. Equivalent results can be obtained for bead positions (see SM\cite{SM}).
To appropriately define a scalar order parameter for the displacive transition, we establish a reference for the positive direction of these displacements.
To do so, we project them onto the eight different degenerate orientations that the $\ce{H_3S}$ molecule can assume in sulfur hydride, and we take the one that maximizes the order parameter $\Delta$, defined as the average of the global displacements over the entire PIMD trajectory made of $N$ steps, i.e., $\Delta = \frac{1}{N} \sum_{j=1}^N \Delta^{(j)}$. For this orientation, we compute also $\Delta_{\textrm{abs}} = \frac{1}{N} \sum_{j=1}^N | \Delta^{(j)}|$.
All  global displacements $\Delta^{(j)}$ belonging to the same PIMD trajectory are computed with respect to the same reference orientation.

To better identify the displacive transition, we exploit a combined analysis of $\Delta$ and $\Delta_{\textrm{abs}}$ (Fig.~\ref{fig:order_parameters}(a)). 
At the transition, the protons freeze in one of the 
degenerate minima of the underlying potential energy surface (PES). When this happens, the order parameter $\Delta$ and its absolute value $\Delta_{\textrm{abs}}$ become indistinguishable, because the fluctuations are suppressed. Simultaneously, the variance of the order parameter, $\sigma^2_\Delta$, suddenly drops 
(Fig.~\ref{fig:order_parameters}(b)). By employing these two criteria, i.e. order parameter saturation and frozen fluctuations, we unambiguously identify the displacive transition line as a function of temperature (Fig.~\ref{fig:phase-diagram}). The critical pressure increases as the temperature is lowered, with a pressure-versus-temperature slope that decreases upon cooling the system. A displacive quantum critical point (QCP) emerges at $T=0$ K, where the transition is purely driven by 
quantum fluctuations.
Here, the PIMD formalism cannot be used to describe the system. However, we can identify the location of the QCP at 
$p_\mrm{QCP} \approx 134 \pm 2$ GPa by extrapolating in temperature and size.

To further characterize the \ce{H_3S} displacive transition, 
we can also look
at the local proton displacement, or local moment
\begin{equation}
    \label{eq:local_order_parameter}
    \Delta^{(j)}_{loc,i} = \vec{r}^{(j)}_{\mrm{H}_i\mrm{S}_{i_1}} \cdot \hat{r}^{(j)}_{\mrm{S}_{i_1}\mrm{S}_{i_2}} - d^{(j)}_{\mrm{S}_{i_1}\mrm{S}_{i_2}}/2,
\end{equation}
defined for the centroid coordinates, as in Eq.~\ref{eq:order_parameter}. Its distribution, gathered for all atoms $H_i$ and PIMD iterations, is shown in Fig.~\ref{fig:order_parameters}(c)
for $T=50$ K and different pressures. In a scenario characterized by a  displacive transition, the H atoms display very different behaviors according to the underlying PES shape. Deep in the centrosymmetric phase, at very high pressures, the potential exhibits a single well and the H atoms stay on average at the S-S segment midpoint, resulting in a unimodal symmetric distribution.Upon  decompression, the potential acquires a displaced global minimum which is eight-fold degenerate. In the distribution of the local parameter $\Delta^{(j)}_{loc,i} $, this is reflected in the development of a bimodal shape, or peak-split density, whose lobes correspond to the positive and negative $\mrm{H}_i$ projections along the corresponding $\mrm{S}_{i_1}$-$\mrm{S}_{i_2}$ direction.
It is the regime of pre-formed local moments, fluctuating across a central barrier.
This corresponds to increased fluctuations of the order parameter, detected by $\sigma^2_\Delta$ 
(Fig.~\ref{fig:order_parameters}(b)), while $\Delta$ is still zero on average. 
Upon further decompression, the potential barrier increases to the point where protons can no longer cross it, freezing in one of the 
degenerate minima; the distribution of the local proton displacement returns unimodal but shifted from the S-S midpoint.
This leads to the creation of a permanent local 
moment, generated by the displaced protons in the R3m phase. Interestingly, although 
$\Delta^{(j)}_{loc,i} $ is a local property, it can be used as an additional probe for the displacive transition, because in \ce{H_3S} the local moments, once frozen, are long-range ordered. 
The other temperatures show the same phenomenology: by reducing the pressure, we observe first a crossover from unimodal to bimodal $\Delta^{(j)}_{loc,i} $ distributions 
with zero order parameter $\Delta$, 
and then an off-centered unimodal shape at the displacive transition (see SM\cite{SM}). 
The density peak splitting associated with the local moment formation does not show a significant temperature dependence.
The displacive transition line approaches the local moment formation pressures only at the lowest temperatures, owing to reduced thermal fluctuations.
At $T=200$ K the displacive critical pressure is 124 $\pm$ 2 GPa, significantly lower than the experimental $T_c$ peak located at about 155 GPa. This mismatch is aligned with other theoretical predictions \cite{Bianco_2018,Taureau2024}.
Moreover, a comparison with classical MD results, also reported in Fig.~\ref{fig:phase-diagram},
shows that the displacive transition pressure is strongly reduced by NQE ($\approx 50$ GPa shift from the classical value at $T = 200$ K).

\begin{figure}[t]
    \centering
    \includegraphics[width=0.95\columnwidth]{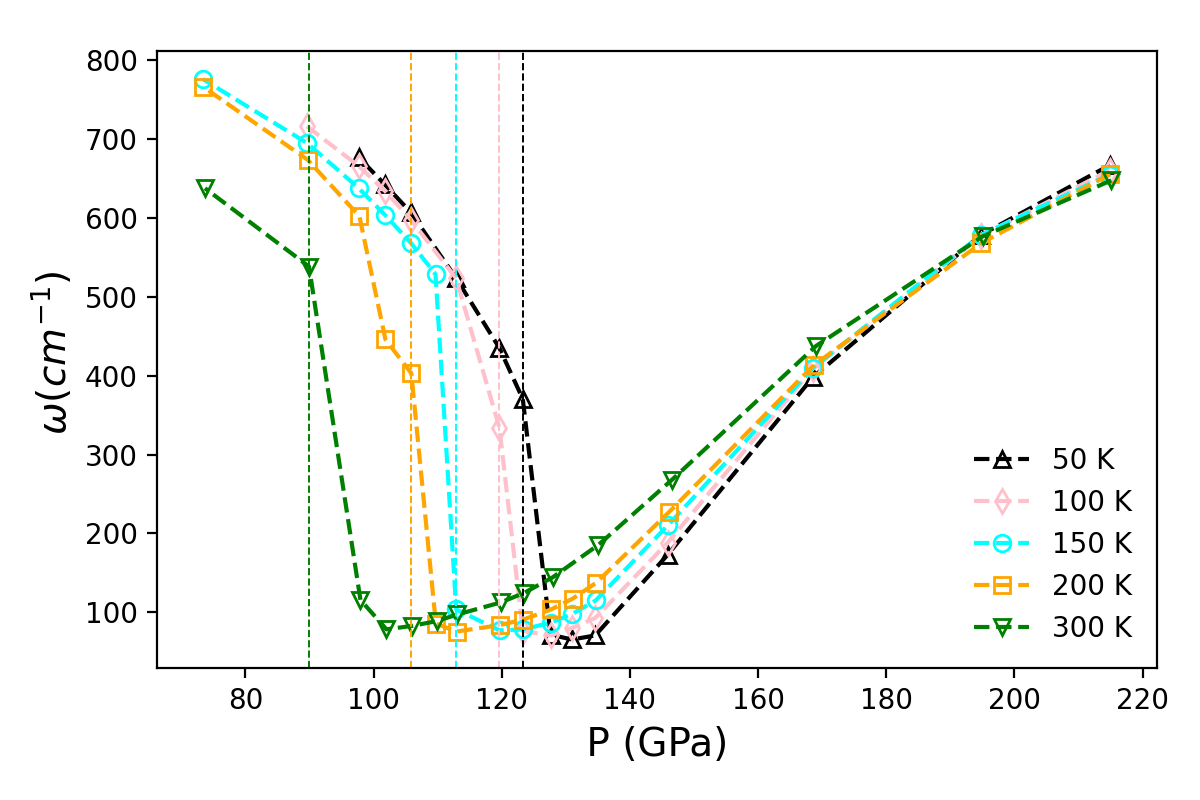}
    \caption{Soft optical modes frequencies as a function of pressure at different temperatures computed for $L=4$
    from PIMD phonons\cite{Morresi2021} (see also the SM\cite{SM}). Colors and symbols are the same as Fig.~\ref{fig:order_parameters}(a)-(b). Dashed vertical lines indicate the displacive transition.}
    \label{fig:phonons_quantum}
\end{figure}

One of the properties most sensitive to structural transitions is lattice dynamics,
encoded
by the imaginary-time phonon Green function:
\begin{equation}
G_{x_i,x_j}(\tau)=-\sqrt{m_i m_j} \langle {\cal T} \delta x_i(\tau) \delta x_j(0) \rangle, 
\label{eq:phonon_green_function}
\end{equation}
with $m_i$ the mass of the $i-$th atom, $\cal{T}$ the time-ordered operator and $\delta x_i(\tau)$ the displacement operator evaluated at the imaginary time $\tau$ along the direction $x$. The quantum correlator $\langle \delta x_i(\tau) \delta x_j(0) \rangle$ is 
accessible by PIMD, where the atomic displacement, $\delta x_i(\tau) =  x_i(\tau) - \langle x_i \rangle$, is defined 
with respect to $\langle x_i \rangle$, the quantum thermal average of the coordinate $x_i$, and evaluated on the $L_\tau$ points in the
imaginary time interval $[0, \beta[$. Within PIMC, one can easily compute the Kubo-transformed version of Eq.~\ref{eq:phonon_green_function}, which corresponds to the static limit of the Matsubara Green function\cite{Morresi2021}. Diagonalizing its inverse yields the anharmonic phonons frequencies in the static approximation\cite{Morresi2021,Morresi2022}.
Dispersive phonon branches are obtained by Fourier transforming the spatial coordinates $x$. 
In Fig.~\ref{fig:phonons_quantum},
we focus on the optical phonons at the $\Gamma$ point that correspond to the modes driving the displacive transition. 
Their patterns are a combination of H shuttling modes along the S-S direction and S-H-S bending modes, leading to the formation of 
distinct \ce{H_3S} molecules in the crystal.
A phonon softening is apparent when approaching the transition from the centrosymmetric phase, while a sharp jump takes place at the transition.
The phonon softening is related to a vanishing curvature of the PES; the sharp frequency increase is due to a stiffer potential experienced by protons once frozen in the polar configuration. Note how at each temperature the frequency jump happens exactly at the corresponding 
critical pressure, as detected from $\Delta$ and its variance $\sigma^2_\Delta$. This is the typical picture of a second-order phase transition, which goes through a sign change of the PES curvature.
The softening of these modes has already been associated to the displacive transition\cite{Bianco_2018,Taureau2024}. 
However, we provide here an unprecedented resolution of their temperature dependence, thanks to the MLIP acceleration and a larger statistics.

\begin{figure}[t]
    \centering
    \includegraphics[width=0.95\columnwidth]{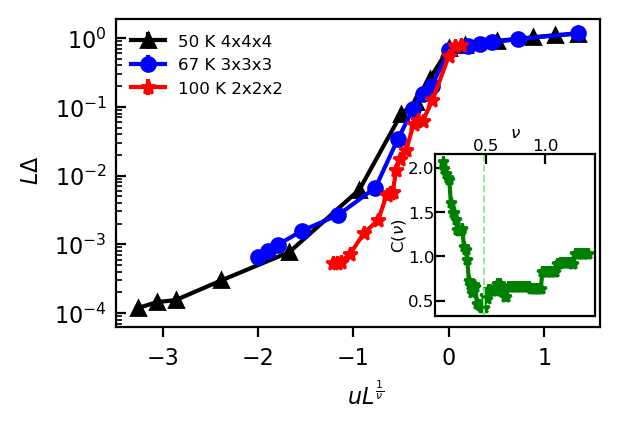}
    \caption{Finite-size scaling of $\Delta$ at $T=50$ K (black triangles), 67 K (blue circles) and 100 K (red stars), and supercell sizes chosen to keep $L_\tau/L$ constant. We use $g[u L^{1/\nu}]$ as a functional form, where $u = \frac{V -V_c}{V_c}$, with $V_c$ the critical volume. Inset: the critical exponent $\nu$ is obtained by minimizing the cost function $C(\nu)$ defined in the text. 
    }
    \label{fig:scaling-analysis}
\end{figure}

We characterize the universality class of the transition at the QCP by performing a finite-size scaling analysis. Close to a QCP, observables $\mathcal{O}$ are expected to scale as \cite{Sachdev2011,Kim2007,Mondaini2023}
\begin{equation}
    \label{eq:scaling_equation}
    \mathcal{O} = L ^{-x_\mathcal{O}} \tilde{\mathcal{O}}(u L^{1/\nu},L_\tau/L^z),
\end{equation}
where $u = \frac{V-V_c}{V_c}$ is the reduced coupling, with $V_c$ the critical volume at a given $L$, and $x_\mathcal{O}$ the scaling dimension of the observable $\mathcal{O}$. $\nu$ is the correlation length exponent, while the dynamical critical exponent $z$ governs the temperature dependence. 
When applying Eq.~\ref{eq:scaling_equation} to the order parameter, i.e. $\mathcal{O} = \Delta$, we assume Lorentz invariance ($z=1$) and vanishing anomalous dimension ($x_\Delta=1$), fulfilled by the Ising universality class\cite{assaad2013pinning,Sachdev2011}. In fact, the computed grid of temperatures and supercell sizes, limited 
by the expensive cost of PIMD simulations, prevents the direct determination of $z$ from our data. Therefore, we assume the validity of the Ising universality class and, in order to check whether our data are compatible with this scenario, we focus on the critical exponent $\nu$. Given $z=1$, we fix the $L_\tau/L$ ratio to be a constant, such that the scaling function in Eq.~\ref{eq:scaling_equation} depends only on one variable. Thus, the scaling analysis is performed using simulations at $T = 50$ K ($L_\tau=100$, $L=4$), 67 K ($L_\tau=75$, $L=3$), and 100 K ($L_\tau=50$, $L=2$) for different volumes.
To obtain $\nu$, we look for the value that makes $L \Delta$ best collapse onto a unique curve (Fig.~\ref{fig:scaling-analysis}). We identify it by minimizing the cost function $C(\nu)=\sum_j|y_{j+1} -y_j|/(\textrm{max}(y_j) -\textrm{min}(y_j))-1$, where $y_j$ are the values of the order parameter for all temperatures, volumes and sizes, ordered according to the $uL^{1/\nu}$ product\cite{vsuntajs2020ergodicity,PhysRevB.104.214201,Mondaini2023}. We found that the optimal value is $\nu\sim 0.483 \pm 0.014$ (see inset of Fig.~\ref{fig:scaling-analysis}), consistent with the 4D Ising universality class ($\nu = \frac{1}{2}$). 
Accordingly, we accurately extrapolated the displacive transition line to the thermodynamic limit, 
by including the logarithmic corrections appearing at the upper critical dimension\cite{brezin1982investigation,aktekin2001finite,kenna2004finite}, such that $p_c \equiv p(V_c)=p_c(\infty) + \alpha L^{-2} \ln(L)^{-1/6}$ at fixed temperature (Fig.~\ref{fig:phase-diagram} and SM\cite{SM}). $p_\mrm{QCP}$ is then obtained by further extrapolating $p_c$ down to zero $T$.

Having established the accurate location of the \ce{H_3S} displacive transition and its 4D Ising universality class, we then quantify the magnitude of quantum fluctuations across the transition. This information is encoded in the imaginary time evolution, accessible in PIMD. Given the importance of the shuttling modes (Fig.~\ref{fig:phonons_quantum}) in driving the transition, it is worth studying the full time dependence of the phonon Green function projected on the proton local displacements (Eq.~\ref{eq:local_order_parameter}), $g(\tau) \equiv G_{ii}(\tau)$, where $i$ is the H atom index. We thus extended the approach of Ref.~\cite{Morresi2021} in order to compute the displacement-displacement correlators in Eq.~\ref{eq:phonon_green_function} by fully retaining their imaginary time dependence. In the case of a local moment formation, $g(\beta/2)$ saturates to a finite value, due to the freezing of local quantum fluctuations, while in absence of preformed local moments $g(\beta/2) \propto \exp(-k \beta) \simeq 0$. Therefore, the variance of $g(\beta/2)$, i.e. $\sigma^2_{g(\beta/2)}$, bears information on the formation of stable local moments. More specifically, we compute the normalized variance $\sigma^2_{g(\beta/2)}/\sigma^2_{g(0)}$, in order to isolate the contribution of retardation effects at $\tau=\beta/2$ from the instantaneous moments formation at $\tau=0$.
Fig.~\ref{fig:order_parameters}(d) shows the pressure dependence of $\sigma^2_{g(\beta/2)}/\sigma^2_{g(0)}$ for all temperatures studied. 
Its maximum at $T = 50$K
matches the pressures where the unimodal-to-bimodal variation occurs in the local density (Fig.~\ref{fig:order_parameters}(c)), while the $\sigma^2_{g(\beta/2)}/\sigma^2_{g(0)}$ ratio decreases sharply in the polar phase, when the H atoms freeze into one of the displaced minima, where fluctuations are strongly suppressed.
At higher temperatures, the $\sigma^2_{g(\beta/2)}/\sigma^2_{g(0)}$ peak is shallower but still well defined, indicating that retardation effects are still relevant, although weaker. Its location is nearly temperature-independent.
To have a more comprehensive view of the 
retardation effects,
in Fig.~\ref{fig:phase-diagram} we plot
$\sigma^2_{g(\beta/2)}/\sigma^2_{g(0)}$ as a heatmap in the $(p,T)$ plane.
It shows how the polar phase is 
characterized by 
suppressed
quantum fluctuations. 
On the other hand, 
large
fluctuations, only marginally irrelevant in 4D, are present in the centrosymmetric side, stretching over a wide 
$(p,T)$ domain.

It is intriguing that the experimental superconducting $T_c$ peak is quite far from the displacive transition line,
while falling in a centrosymmetric region with strong 
retardation effects
spawned by the QCP. 
It is already known that NQE can enhance the superconducting critical temperature of a BCS superconductor, as shown in the atomic phase of pristine hydrogen\cite{dangic2024large}. This requires going beyond the standard Migdal-Eliashberg framework\cite{Marsiglio2020} by taking into account phonon renormalization effects\cite{Marsiglio1991,Esterlis2018} through the inclusion of the full frequency dependence of the phonon Green function, beside the multi-band and the $k$-dependence nature of the electron-phonon interaction\cite{sano2016effect,Lucrezi2024}, and its vertex corrections\cite{pietronero1995nonadiabatic,Grimaldi1995,Mishra2025}. In the case of $\ce{H_3S}$, source of these effects is 
the proximity of the displacive QCP, which then calls for a more appropriate evaluation of the superconducting $T_c$ in this material. 
At the same time, it would be desirable to estimate the electron-phonon coupling within a non-perturbative framework, such as the one proposed in Ref.~\cite{bianco2023non}. Indeed, the presence of a displacive QCP has been invoked to explain the enhancement of superconductivity in some dilute metals like $\ce{SrTiO_3}$ \cite{Ngai1974,vanderMarel2019,Collignon2019,gastiasoro2020superconductivity,Kiselov2021,Volkov2022}, by relying on an electron-pairing mechanism beyond linear order in the electron-phonon vertex, namely mediated by two optical phonons\cite{Ngai1974}. Although it remains to verify whether $\ce{H_3S}$ meets similar
conditions, the application of the latter picture to this material is tempting, owing to the presence of optical phonons with strong spectral weight\cite{Capitani2017}, significantly softened at the displacive transition (Fig.~\ref{fig:phonons_quantum}), and to the strong renormalization of both one-body and two-body phonon Green functions (Fig.~\ref{fig:phase-diagram} and SM\cite{SM}). 

To conclude, we have characterized the phase diagram of $\ce{H_3S}$ over a wide range of pressures and temperatures, by unveiling the 4D Ising universality class of its displacive quantum criticality, the location of the QCP and its transition line from BLYP-MLIP driven PIMD. The emerging scenario for the high-$T_c$ superconductivity in $\ce{H_3S}$ appears more complex than previously thought. Indeed, according to our calculations the maximum of the experimental superconducting $T_c$ falls in a 
centrosymmetric region dominated by nuclear quantum fluctuations and retardation effects surrounding the QCP, which could both affect superconductivity.

\textit{Acknowledgments} - We acknowledge GENCI for providing computational resources on the IDRIS Jean-Zay supercomputing clusters and TGCC Joliot-Curie Rome partition under project number A0190906493. We are grateful for computational resources from EuroHPC for the computational grant EHPC-EXT-2024E01-064 allocated on Leonardo (booster partition). We acknowledge EPICURE, a EuroHPC Joint Undertaking initiative for supporting this project on the booster partition of Leonardo through the EuroHPC JU 2024E01 call for proposals for extreme scale access mode. We thank the European High Performance Computing Joint Undertaking (JU) for the support through the "EU-Japan Alliance in HPC" HANAMI project (Hpc AlliaNce for Applications and supercoMputing Innovation: the Europe - Japan collaboration). A.R. acknowledges financial support from JST BOOST (Grant No.~JPMJBY24F3). We thank Pr. Ryotaro Arita, Pr. Matteo Calandra, Dr. Lorenzo Monacelli, Dr. Tommaso Morresi, and Dr. Kosuke Nakano for useful discussions.

\textit{Data availability} - The data that support the findings of this article are openly available at \cite{zenodo}.

\nocite{Ceriotti2010,Giannozzi_2009,Giannozzi_2017,Gilmer,Bronstein2021}

\bibliography{main}

\clearpage
\pagestyle{empty}
\foreach \i in {1,...,11} {
    \begin{figure}[p]
        \centering
        \includegraphics[page=\i, width=\textwidth]{supplemental_material.pdf }
    \end{figure}
    \clearpage
}

\end{document}